\documentclass[twoside,twocolumn,9pt]{article}
\usepackage{extsizes}
\usepackage[super,sort&compress,comma]{natbib} 
\usepackage[version=3]{mhchem}
\usepackage[left=1.5cm, right=1.5cm, top=1.785cm, bottom=2.0cm]{geometry}
\usepackage{balance}
\usepackage{times,mathptmx}
\usepackage{sectsty}
\usepackage{graphicx} 
\usepackage{lastpage}
\usepackage[format=plain,justification=justified,singlelinecheck=false,font={stretch=1.125,small,sf},labelfont=bf,labelsep=space]{caption}
\usepackage{float}
\usepackage{fancyhdr}
\usepackage{fnpos}
\usepackage[english]{babel}
\usepackage{array}
\usepackage{droidsans}
\usepackage{charter}
\usepackage[T1]{fontenc}
\usepackage[usenames,dvipsnames]{xcolor}
\usepackage{setspace}
\usepackage[compact]{titlesec}
\usepackage{hyperref}
\newcommand{\ue}{\mathrm{e}}
\newcommand{\ud}{\mathrm{d}}
\newcommand{\onlinecite}[1]{\hspace{-1 ex} \nocite{#1}\citenum{#1}} 
\usepackage {amssymb}
\usepackage {latexsym}
\usepackage{epstopdf}

\definecolor{cream}{RGB}{222,217,201}

\begin{document}

\thispagestyle{plain}



\renewcommand{\thefootnote}{\fnsymbol{footnote}}
\renewcommand\footnoterule{\vspace*{1pt}%
\color{cream}\hrule width 3.5in height 0.4pt \color{black}\vspace*{5pt}} 
\setcounter{secnumdepth}{5}

\makeatletter 
\renewcommand\@biblabel[1]{#1}            
\renewcommand\@makefntext[1]%
{\noindent\makebox[0pt][r]{\@thefnmark\,}#1}
\makeatother 
\renewcommand{\figurename}{\small{Fig.}~}
\sectionfont{\sffamily\Large}
\subsectionfont{\normalsize}
\subsubsectionfont{\bf}
\setstretch{1.125} 
\setlength{\skip\footins}{0.8cm}
\setlength{\footnotesep}{0.25cm}
\setlength{\jot}{10pt}
\titlespacing*{\section}{0pt}{4pt}{4pt}
\titlespacing*{\subsection}{0pt}{15pt}{1pt}


\makeatletter 
\newlength{\figrulesep} 
\setlength{\figrulesep}{0.5\textfloatsep} 

\newcommand{\topfigrule}{\vspace*{-1pt}%
\noindent{\color{cream}\rule[-\figrulesep]{\columnwidth}{1.5pt}} }

\newcommand{\botfigrule}{\vspace*{-2pt}%
\noindent{\color{cream}\rule[\figrulesep]{\columnwidth}{1.5pt}} }

\newcommand{\dblfigrule}{\vspace*{-1pt}%
\noindent{\color{cream}\rule[-\figrulesep]{\textwidth}{1.5pt}} }

\makeatother

\twocolumn [
  \begin{@twocolumnfalse}
\vspace{3cm}
\sffamily
\begin{tabular}{m{-0.5cm} p{18cm} }

 & \noindent\huge{\textbf{Suppression of vacancies boosts thermoelectric performance in type-I clathrates}} \\
\vspace{0.3cm} & \vspace{0.3cm} \\

 & \noindent\large{Xinlin Yan,$^{\ast}$\textit{$^{a}$} Matthias Ikeda,\textit{$^{a}$} Long Zhang,\textit{$^{b}$} Ernst Bauer,\textit{$^{a,f}$}  Peter Rogl,\textit{$^{c}$} Gerald Giester,\textit{$^{d}$} Andrey Prokofiev,\textit{$^{a}$} and Silke Paschen$^{\ast}$\textit{$^{a}$} } \\

\vspace{0.3cm} & \vspace{0.3cm} \\

 & \noindent\normalsize{Intermetallic type-I clathrates continue to attract attention as promising thermoelectric materials. Here we present structural and thermoelectric properties of single crystalline Ba$_8$(Cu,Ga,Ge,$\Box$)$_{46}$, where $\Box$ denotes a vacancy. By single crystal X-ray diffraction on crystals without Ga we find clear evidence for the presence of vacancies at the 6$c$ site in the structure. With increasing Ga content, vacancies are successively filled. This increases the charge carrier mobility strongly, even within a small range of Ga substitution, leading to reduced electrical resistivity and enhanced thermoelectric performance. The 
largest figure of merit $ZT=0.9$ at 900 K is found for a single crystal of approximate composition Ba$_8$Cu$_{4.6}$Ga$_{1.0}$Ge$_{40.4}$. This value, that may further increase at higher temperatures, is one of the largest to date found in transition metal element-based clathrates.

} \\

 \end{tabular}

  \end{@twocolumnfalse} \vspace{0.6cm}

  ]

\renewcommand*\rmdefault{bch}\normalfont\upshape
\rmfamily
\section*{}
\vspace{-1cm}


\footnotetext{\textit{$^{a}$~Institute of Solid State Physics, Vienna University of Technology, Wiedner Hauptstr. 8--10, 1040 Vienna, Austria. Fax: +43 1 58801 13899; Tel: +43 1 58801 13716; E-mail: paschen@ifp.tuwien.ac.at; yan@ifp.tuwien.ac.at}}
\footnotetext{\textit{$^{b}$~State Key Laboratory of Metastable Materials Science and Technology, Yanshan University, Qinhuangdao, Hebei 066004, China }}
\footnotetext{\textit{$^{c}$~Institute of Materials Chemistry and Research, Vienna University, W\"ahringerstr. 42, 1090 Vienna, Austria }}
\footnotetext{\textit{$^{d}$~Institute of Mineralogy and Crystallography, University of Vienna, Althanstr. 14, 1090 Vienna, Austria }}
\footnotetext{\textit{$^{f}$~Christian Doppler Laboratory for Thermoelectricity, TU Wien, WiednerVienna, Austria }}




\section{Introduction} \label{intr}
Intermetallic type-I clathrates are promising materials for high-temperature thermoelectric (TE) applications. The unique TE properties of these materials are associated with the crystal structure which is composed of polyhedral cages formed by covalently bonded host framework atoms and  guest atoms ionically bonded in these cages. The guest atoms can act as rattlers,\cite{95Sla,97Sla,99Coh.No,00Nol.38,00Nol2.Ch,01Nol.Sl} creating low-lying optical modes. If the frequency of the rattling modes lies within the acoustic range, an interaction of acoustic and optical modes may result and lead to low lattice thermal conductivities.\cite{08Chr.Ab,12Euc.Pa,14Pai.Eu,2016XShi,2016She,2016Bha2,2017Bha,2017Rob} The charge transport is mainly governed by the framework,\cite{95Sla,97Sla} giving rise to comparably high charge carrier mobilities. The combination of these properties is beneficial for reaching high values of the dimensionless thermoelectric figure of merit $ZT=TS^2/(\rho \kappa$), where $T$ is the absolute temperature, $S$ the Seebeck coefficient, $\rho$ the electrical resistivity, and $\kappa$ the thermal conductivity. $\kappa$ is usually composed of the lattice thermal conductivity $\kappa_{\rm ph}$ and the electronic thermal conductivity $\kappa_{\rm e}$. As a common parameter of $S$, $\rho$, and $\kappa_{\rm e}$, the charge carrier concentration plays a crucial role in determining $ZT$. Generally, type-I clathrates can be regarded as Zintl compounds:\cite{39Zin,11She.An,2016She} The guest atoms (for anionic clathrates) donate their valence electrons to the framework atoms which use them in covalent framework bonds. If all valence electrons are used up, the system is an insulator. If there are more (less) valence electrons than needed to complete the bonding, the system is an n-type (a p-type) semiconductor. Thus, the composition as well as details of the crystal structure (e.g., whether there are vacancies and how atoms distribute in the crystal structure\cite{2016Les,2016Ang,2015Bob}) are critical for the thermoelectric properties of clathrates.\cite{10Chr.Jo,11She.An,2016She} So far, the type-I clathrates most promising for thermoelectric applications are Ba-Ga-Ge(Si)-based compounds with a Ga content around 16 atoms per unit cell (u.c.). Transition metal (TM) element  containing clathrates also have been widely studied\cite{2016She} and remarkable $ZT$ values have been reported for clathrates such as the Ba-Au-Ge system\cite{11Zha.Bo} and the Ba-Zn-Ge-Sn system.\cite{2013Fal} Cu-containing clathrates, interesting for the low price of Cu, still have low $ZT$ values due to the non-optimized charge carrier concentration and low charge carrier mobility. For Ba$_8$Cu$_x$Ge$_{46-x}$ clathrates, studies showed that vacancies exist when $x\lesssim 5.5$.\cite{09Mel.An} Previous studies have suggested that it is unfavorable if the clathrate optimized for charge carrier concentration contains vacancies because these may scatter charge carriers and reduce the charge carrier mobility.\cite{2016She} Attempts were then performed to change the atomic environment by elemental substitution, for instance with Sn (Ref.\onlinecite{2015Xu}) or Ga.\cite{05Ho.An,2016Les} Interesting results including improved charge carrier mobility and enhanced TE properties have been observed.\cite{2015Xu,05Ho.An}

\begin{figure}[ht]
 \centering
 \includegraphics[width=0.5\textwidth]{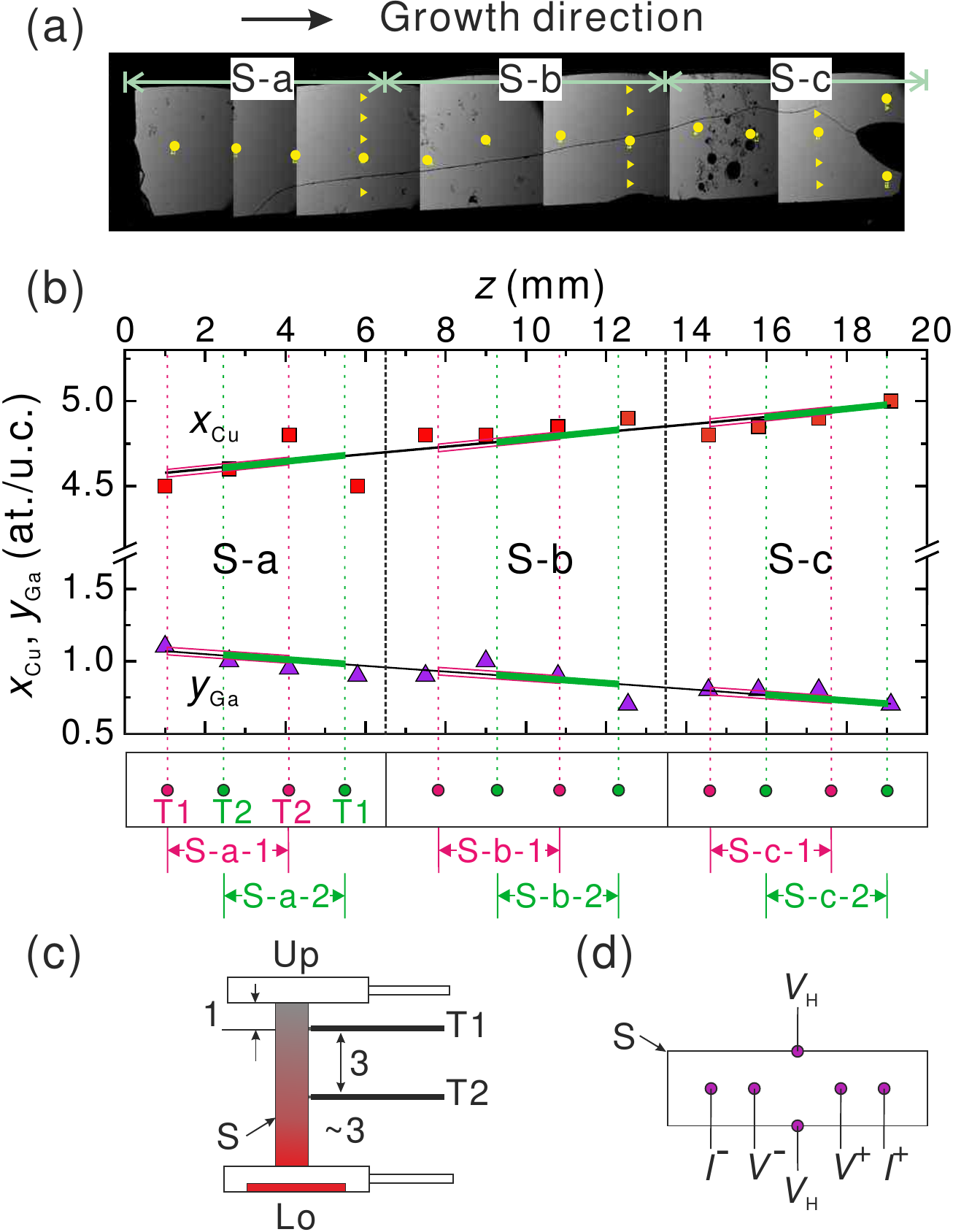}
 \caption{ (Color online) (a) SEM image of the as-growth single crystal. The yellow points represent the locations for composition measurements by EDX. We define the crystal coordinate $z$ along the crystal growth direction. (b) Cu and Ga contents vs $z$. Dashed lines are cutting lines for samples S-a, S-b, and S-c. (c) Sketch of the ULVAC system's sample holder (Up and Lo: upper and lower electrodes; length in mm). The asymmetric arrangement of the temperature sensors labeled T1 and T2 around the center of the sample (S) can be seen. Two measurements (i = 1 and 2) for each part were then performed. The locations of T1 and T2 in each measurement are indicated in (b) bottom. (d) Sketch of wire connections on the sample (S) for PPMS measurements.}
 \label{Comp.Mic}
\end{figure}

In the present work, we study variations of the Ga content in the framework of Ba$_8$(Cu,Ga,Ge,$\Box$)$_{46}$ clathrates, and their effects on the structural and TE properties.  
For this purpose, we grew two different large single crystals by the floating-zone technique, one with and the other without Ga. The Ga-containing 
as-grown crystal shows compositional gradients as seen in other crystals prepared by the floating-zone technique. \cite{06Sar.Sv,09Hou.Zh,11Mug.Na,12Nag.Mu,2014Pro,06Cai.Zh,09Chr.Jo} 
This can be exploited to study the compositional dependence of the TE properties. Interestingly, in a very narrow composition range, the charge carrier mobility is sizably enhanced with increasing Ga content, leading to reduced electrical resistivity and improved TE performance. By comparing the crystal structure of a crystal with approximate composition Ba$_8$Cu$_{4.8}$Ga$_1$Ge$_{40.2}$ with that of a Ga-free crystal of approximate composition Ba$_8$Cu$_{4.8}$Ge$_{41.2}$, we confirm that vacancies exist in the latter crystal and Ga atoms fill these vacancies. The vacancy filling by Ga substitution is responsible for the enhanced TE performance.

\begin{figure}[ht]
 \includegraphics[width=0.5\textwidth]{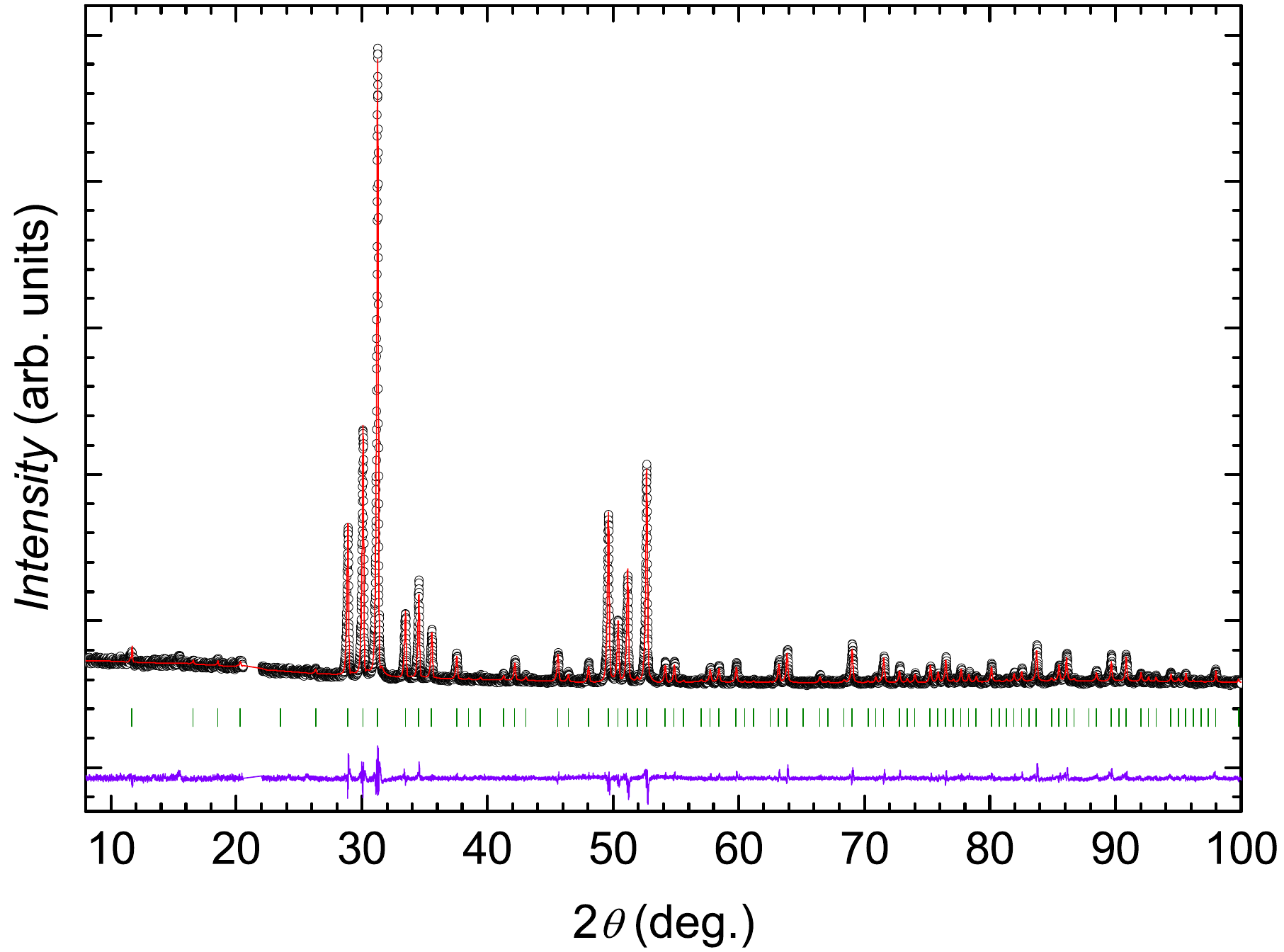}%
 \caption{\label{Riet-diff} (Color online) Rietveld refinement (red line) of the XPD data (black circles) of S-a. 
}
 \end{figure}

\section {Experimental section}
\subsection {Crystal growth}\label{Cg}
As starting materials for the growth of the Ga-containing crystal, two cylindrical rods with the same nominal composition Ba$_8$Cu$_{4.8}$Ga$_1$Ge$_{40.2}$ were prepared in a high-frequency induction furnace from high-purity elements. One rod with 7\,mm in diameter and 60\,mm in length served as the feed rod, the other one with the same diameter and 20\,mm in length as the seed for the crystal growth. 

The crystal was grown in a 4-mirror furnace equipped with 1000\,W halogen lamps. The pulling speed of the rod was 3-5\,mm/h. Both rods rotated oppositely (speed: $\sim$8\, rpm) to ensure efficient mixing of the liquid and an uniform temperature distribution in the molten zone. A pressure of 1.5 bar of Ar was used during the crystal growth. For more details on the growth conditions, please refer to our previous work.\cite{2014Pro}   

To elucidate the effects of Ga substitution, a Ga-free single crystal with a nominal composition Ba$_8$Cu$_{4.8}$Ge$_{41.2}$ was grown using the same synthesis processes. 

\begin{table*}
\small
  \caption{\ Average compositions derived from Eq.\,\ref{eqCon1} for different samples/parts (Fig.\,\ref{Comp.Mic}), theoretical carrier concentration $n$ calculated from Eq.\,\ref{Zintl}, experimental charge carrier concentration $n_{\rm H}$ and mobility $\mu_{\rm H}$ (both at 300 K) evaluated from Hall effect measurements, electrical resistivity $\rho(300$ K), Seebeck coefficient $S(300$ K), and effective mass $m^*$ (300 K) derived from Eq.\,\ref{eqS1}.}
  \label{Tab:tab1}
  \begin{tabular*}{\textwidth}{@{\extracolsep{\fill}}l c c c c c c c}
    \hline
    Code &  Composition  & $n$ (-e/u.c.) &$n_{\rm H}$ (-e/u.c.) & $\rho$ ($\mu \Omega$ cm) & $S$ ($\mu$ V/K) & $m^*$/m$_{\rm e}$ & $\mu_{\rm H}$ (cm$^2$/Vs)  \\
    \hline
 S-a-1 & Ba$_8$Cu$_{4.61}$Ga$_{1.04}$Ge$_{40.35}$ & 1.13 &  & 470 & -37.4 &  &  \\
S-a   & Ba$_8$Cu$_{4.63}$Ga$_{1.02}$Ge$_{40.35}$ & 1.09 & 0.72 & 488 & -37.0 & 1.29 & 21.2 \\
S-a-2 & Ba$_8$Cu$_{4.64}$Ga$_{1.01}$Ge$_{40.35}$ & 1.06 &  & 506 & -36.5 &  &  \\
S-b-1 & Ba$_8$Cu$_{4.77}$Ga$_{0.89}$Ge$_{40.33}$ & 0.79 &  & 550 & -42.1 &  &  \\
S-b   & Ba$_8$Cu$_{4.78}$Ga$_{0.89}$Ge$_{40.33}$ & 0.77 & 0.67 & 563 & -40.4 & 1.35 & 19.3 \\
S-b-2 & Ba$_8$Cu$_{4.79}$Ga$_{0.87}$Ge$_{40.33}$ & 0.75 &  & 576 & -38.7 &  &  \\
S-c-1 & Ba$_8$Cu$_{4.91}$Ga$_{0.77}$Ge$_{40.32}$ & 0.50 &  & 737 & -35.8 &  &  \\
S-c   & Ba$_8$Cu$_{4.92}$Ga$_{0.75}$Ge$_{40.33}$ & 0.47 & 0.62 & 781 & -42.9 & 1.36 & 15.1 \\
S-c-2 & Ba$_8$Cu$_{4.94}$Ga$_{0.74}$Ge$_{40.32}$ & 0.44 &  & 826 & -50.0 &  &  \\
    \hline
  \end{tabular*}
\end{table*}

\subsection {Characterization}

Single crystals with a size of about 60 $\mu$m were mechanically isolated from crushed single crystal pieces. Inspection on an AXS-GADDS texture goniometer assured high crystal quality, and provided unit cell dimensions and Laue symmetry of the specimens prior to an X-ray intensity data collection on a four-circle Nonius Kappa diffractometer equipped with a CCD area detector employing graphite monochromated Mo-K$\alpha$ radiation ($\lambda$ = 0.071069 nm) at 300 K. The orientation matrix and unit cell parameters were derived using the program DENZO.\cite{1998Non} No absorption corrections were necessary because of the rather regular crystal shapes and the small dimensions of the investigated specimens. The structures were solved by direct methods and refined with the Oscail program. A quantitative analysis of the structural details was done with the program SHELXS-97.\cite{1997She}  

X-ray powder diffraction (XPD) data were collected using a HUBER-Guinier image plate system (Cu K$_{\alpha_1}$, $8^\circ \leq 2\theta \leq 100^\circ$). Lattice parameters were calculated by least squares fits to indexed $2\theta$ values
employing Ge ($a_{\rm {Ge}}=0.5657906$\,nm) as internal standard. Rietveld refinements were performed for the XPD data by using the program FULLPROF.\cite{90Ro.Ca}

The composition was determined by energy dispersive x-ray spectroscopy (EDX) in a scanning electron microscope (SEM)
operated at 20\,kV (Zeiss Supra 55VP, probe size: $1 \mu$m). The measured compositions were normalized to 8 Ba atoms per unit cell (at./u.c.) with an assumption of no vacancy in the framework.

\subsection {Physical properties}\label{phymess}

The electrical resistivity $\rho$ and Seebeck coefficient $S$ were measured with a ZEM-3 (ULVAC-Riko, Japan) between room temperature and $600^\circ\mathrm{C}$. In order to fulfill the size limitation for the measurements, the as growth single crystal was cut into 3 parts with $\sim$7 mm in length (samples S-a, S-b, and S-c shown in Fig.\,\ref{Comp.Mic}).

The two temperature sensors (T1 and T2) are asymmetrically arranged around the sample center (see Fig.\,\ref{Comp.Mic} (c)). To maximize the number of different measurement geometries and thus the amount of data from different sample compositions, each sample was measured along two directions, as indicated in Fig.\,\ref{Comp.Mic} (b) bottom. Both $\rho$ and $S$ have uncertainties of $<5\%$.

The thermal conductivity at high temperatures (300--900 K) was derived from the thermal diffusivity $D_t$ measured using the flash method with a Flashline-3000 (ANTER, USA), the specific heat $C_p$ estimated using the Dulong-Petit approximation, and the density $D$, using the relation $\kappa = D_t C_p D$. A disc-like sample (diameter $\phi = 6$ mm, thickness $t=1$ mm), selected from near the beginning of the as-grown single crystal, was used.

Hall effect measurements were performed in a physical property measurement system (PPMS, Quantum Design, Model 6000) in the temperature range 2 to 300\,K, in magnetic fields up to 9\,T. We used a standard 6-point ac technique in which the Hall contacts are perpendicular to both the magnetic field and the electrical current. Small longitudinal resistivity components due to contact misalignment were subtracted by magnetic field reversal. At selected temperatures we confirmed that the Hall response is linear in field. The charge carrier concentration $n_{\rm H}$ was calculated using a simple one-band model $n_{\rm H}=1/({\rm e} R_{\rm H})$. The Hall mobility $\mu_{\rm H}$ was determined by $\mu_{\rm H}=R_{\rm H}/\rho$. To determine the average charge carrier concentration of each part, the Hall contacts were positioned around the center of each part (Fig.\,\ref{Comp.Mic} (d)).

Specific heat measurements under zero magnetic field were performed with a PPMS by a standard relaxation method between 2 and 300\,K.

\begin{figure}[H]
 \includegraphics[width=0.5\textwidth]{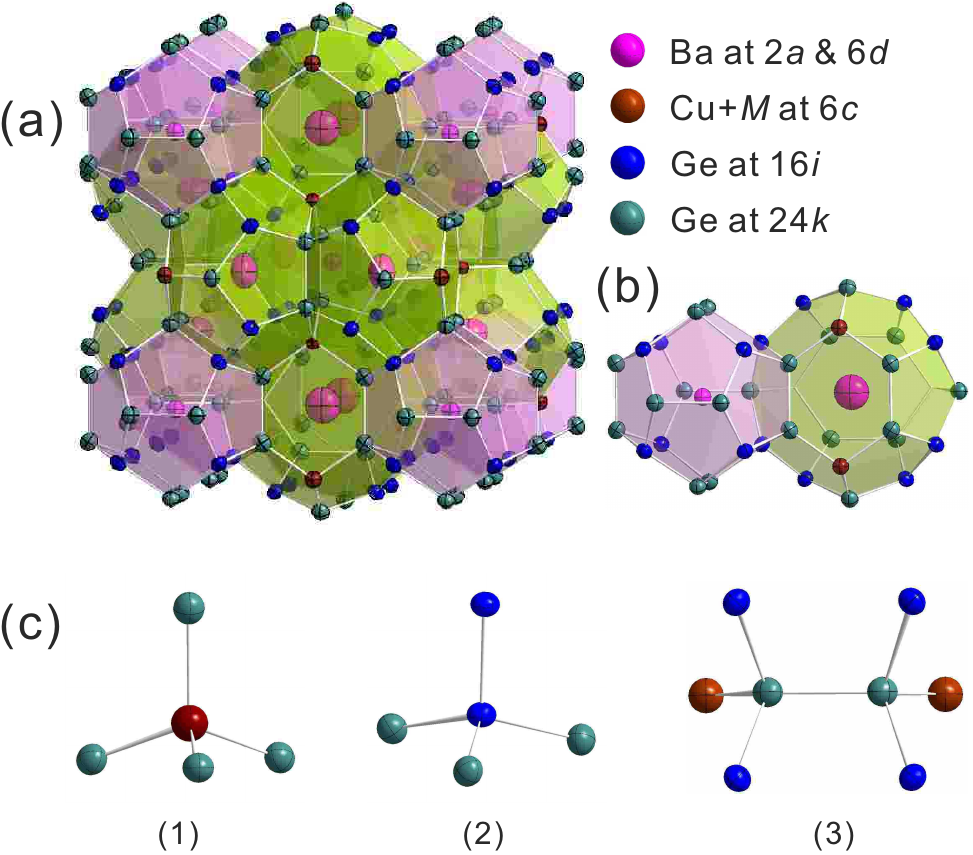}%
 \caption{\label{Crytal} (Color online) (a) Crystal structure of type-I clathrates. Anisotropic thermal atomic displacements, shown as distorted spheres, refer to the single crystal
diffraction refinements; (b) two adjacent cages; (c) the environments for atoms at the $6c$ (1), 16$i$ (2), and 24$k$ (3) site. 
}
 \end{figure}

\section {Results and discussion}
\subsection {Chemical properties of as-grown single crystals}\label{chemical}

Both the Ga-containing and the Ga-free as-grown single crystals have a length of $\sim$22 mm and a diameter of $\sim$7 mm. XPD, optical microscopy, and SEM measurements confirmed the type-I clathrate structure (no. 223, $Pm\bar{3}n$) and the high quality with no visible foreign phases. Figure\,\ref{Riet-diff} shows an example from XPD. The crystals are stable in air and mechanically strong. Lattice parameters from different parts in each crystal are very similar. The average value is 1.06975(2)\,nm for our single crystal of nominal composition Ba$_8$Cu$_{4.8}$Ga$_1$Ge$_{40.2}$, close to values found for polycrystalline samples of similar composition, e.g., 1.0696(1)\,nm for Ba$_8$Cu$_{5.25}$Ga$_1$Ge$_{39.67}$ Ref.\onlinecite{05Ho.An} and 1.0702 nm for Ba$_8$Cu$_{5}$Ga$_1$Ge$_{40}$.\cite{2016Les} Also the average lattice parameter 1.06928(2)\, nm for our single crystal of nominal composition Ba$_8$Cu$_{4.8}$Ge$_{41.2}$ is close to the literature values for polycrystals of similar compositions.\cite{09Mel.An,06Joh.Si,2014Chen,2016Les}

\begin{figure}[H]
 \includegraphics[width=0.5\textwidth]{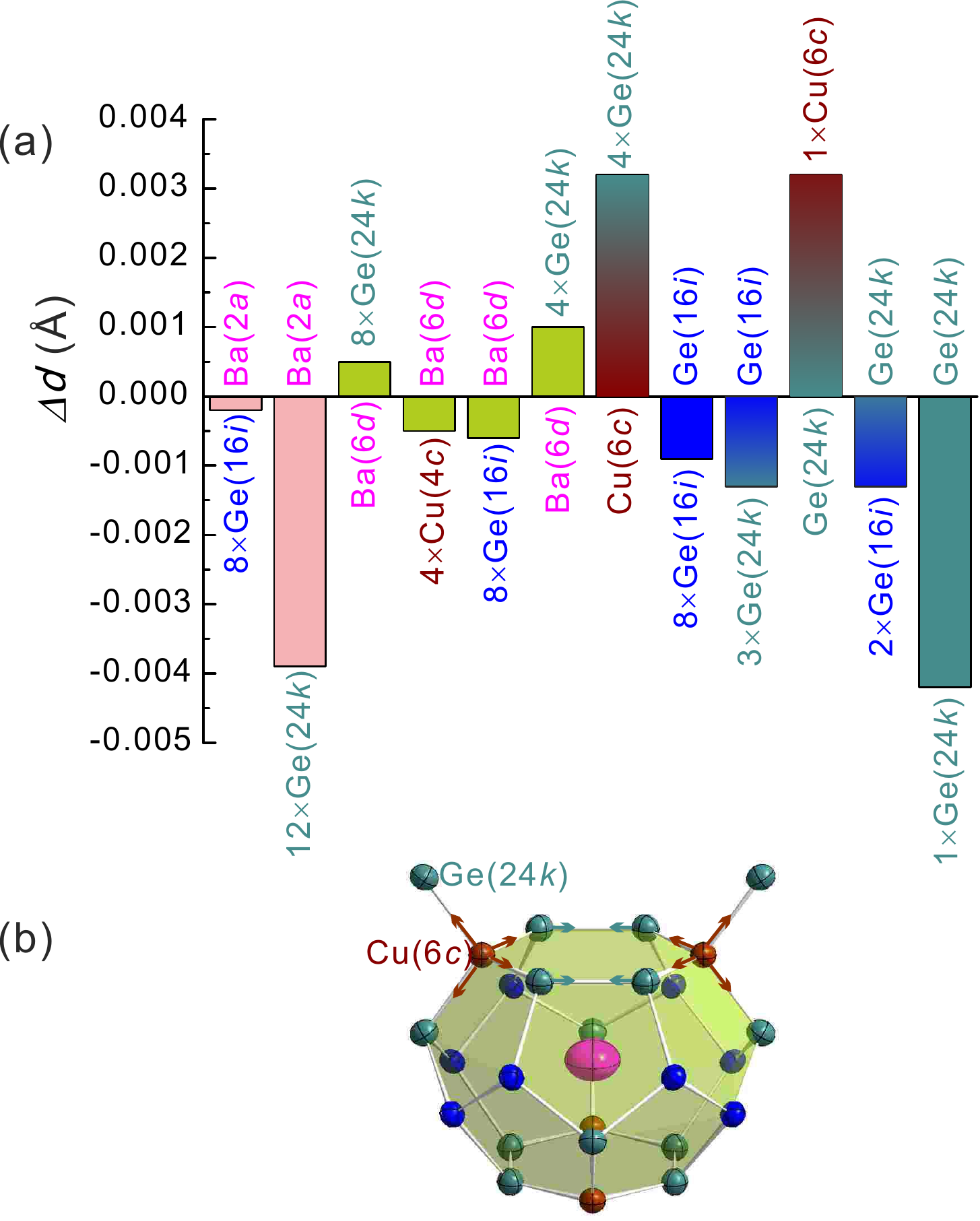}%
 \caption{\label{distance0} (Color online) (a) Distance changes induced by Ga substitution for the respective centered atoms of Fig.\,\ref{Crytal} (b) and (c)); (b) Sketch showing the elongation of the interatomic distance Cu(6$c$)-Ge(24$k$) and the shrinkage of Ge(24$k$)-Ge(24$k$) bond. 
}
 \end{figure}

The composition determination by EDX was performed along both the growth direction and the radial direction (Fig.\,\ref{Comp.Mic} (a)). In the Ga-free crystal, the composition differences in both directions are very small. The average composition is Ba$_8$Cu$_{5.0}$Ge$_{41.0}$, with a Cu content slightly above that of the nominal composition Ba$_8$Cu$_{4.8}$Ge$_{41.2}$. The composition of the Ga-containing crystal, however, changes distinctly along the growth direction (Fig.\,\ref{Comp.Mic} (b)), indicating a complex reaction scheme during the crystal growth process.\cite{2014Pro} There is a clear correlation between the Cu and the Ga content, with the Cu content increasing and the Ga content decreasing along the growth direction. This is the behavior expected within an electron-balanced scheme of the Zintl rule.\cite{39Zin,11She.An,2016She} The changes of the Cu and Ga contents along the coordinate $z$ (defined along the growth direction, see Fig.\,\ref{Comp.Mic} (b)) are described by  
\[x_{\rm Cu}(z) = 0.022z+4.556\]
and
\begin{equation}
  y_{\rm Ga}(z) = -0.020z+1.089\,\hspace{0.2cm}  .
\label{eqCon1}
\end{equation}
\noindent The average composition of each piece S-a to S-c used for physical property investigations can then be estimated from these relations.

\subsection {Structure analysis of single crystals}\label{Cryt}

Due to their similar X-ray scattering factors, the occupations of Cu, Ge, and Ga in Ba$_8$Cu$_{4.8}$Ga$_1$Ge$_{40.2}$ are hard to derive from X-ray diffraction. Thus a number of different techniques\cite{2007Chr,2016Les,09Mel.An,2017Rob,2015Xu,2016Ang,2015Bob,2014He,2016She,2014Chen} have been applied to determine structural details and several basic rules have been established. For instance, TM elements and possible vacancies preferentially occupy the 6$c$ site \cite{09Mel.An,11Yan.Gr,13Fal.Gry,12Yan.Ch,13Yan.Ba,2016Les,2015Xu,2014Chen,11Zha.Bo,06Joh.Si} and avoid direct 13 element bonding.\cite{2007Chr,2012Rou,2015Bob,2015Sui,2016Ang,2017Rob} Structural details can also be extracted from thermal parameters obtained by refining diffraction data or NMR measurements,\cite{2014Chen,2015Sir} or from changes of interatomic distances induced by composition variations as shown here.

The refinements of the single crystal data for both crystals revealed isomorphism of the type-I clathrate structure (SG: 223, $Pm\bar 3n$). The heavier Ba atoms are located at the $2a$ and $6d$ sites, and the framework sites are $6c$, $16i$, and $24k$. For further refinements, we first assumed an ordered model for the framework, i.e., Cu fully located at the $6c$ site and Ge/Ga at the 16$i$ and 24$k$ sites. As the differences between Cu, Ge, and Ga atoms are essentially invisible in X-ray diffraction, this is certainly a plausible way forward. The refinements gave very good reliability factors and reasonable thermal parameters (temperature factors). Though vacancies have been evidenced in some ternary Ba$_8$Cu$_x$Ge$_{46-x}$ clathrates,\cite{09Mel.An,06Joh.Si} we could not pin down their existence in our refinements. This might be due to a very low level of vacancies in our crystals. 
The almost spherical shapes of the atoms at the $6c$ site (Fig.\,\ref{Crytal}), that reflect the thermal parameters of the refinement, make it difficult to recognize vacancies in the structure. Even a model with a site splitting at the $24k$ site does not reveal any sizable distortion from a spherical shape. A change of the occupation in the structure model for the refinement does not change the interatomic distances in the structure. Thus, a comparison of interatomic distances in our two crystals may be the most sensitive means to reveal vacancies. We focused on the following distance changes induced by Ga substitution: The distance between Ba atoms and the framework, and the distance between coordinated framework atoms (see Fig.\,\ref{Crytal} (b) and (c, 1-3)). The results are given in Table\,\ref{StruDat-d} and visualized in Fig.\,\ref{distance0} (a). The Ga substitution of about 1 at./u.c. shrinks the small cages by shortening the interatomic distances Ba(2$a$)-Ge(24$k$), but leaves the large cages essentially unchanged. All large changes shown in Fig.\,\ref{distance0} are associated with the atoms at the 24$k$ site, giving a first glance that Ga could replace Ge at the $24k$ site in Ba$_8$Cu$_{4.8}$Ga$_1$Ge$_{40.2}$ just as Sn does in Ba$_{8.0}$Cu$_{5.1}$Sn$_{0.7}$Ge$_{40.2}$.\cite{2015Xu} However, the structure is more complex here because locating Ga at the $24k$ site alone cannot explain (1) the shrinkage of the interatomic distance Ge(24$k$)-Ge(24$k$), which should be elongated due to the slightly larger covalent radii of Ga compared to that of Ge; (2) the elongation of the interatomic distance Cu(6$c$)-Ge(24$k$); and (3) the shrinkage of the small cages. We therefore introduce a vacancy-filling model in which Ga atoms fill vacancies at the 6$c$ site, leading to an increased interatomic distance Cu(6$c$)-Ge(24$k$) and a shortened distance Ge(24$k$)-Ge(24$k$) as sketched in Fig.\,\ref{distance0} (b). This strongly suggests that vacancies exist in Ba$_8$Cu$_{4.8}$Ge$_{41.2}$ and are filled by atoms induced by the Ga substitution. A possible structure model for Ba$_8$Cu$_{4.8}$Ge$_{41.2}$, with Cu+Ge (the Cu content is fixed to 5.0 at./u.c. from EDX, M1 in Table\,\ref{StruDat}) atoms occupying the $6c$ site, and a model for Ba$_8$Cu$_{4.8}$Ga$_1$Ge$_{40.2}$, with Cu+Ge (the Cu content is fixed to 4.6 at./u.c. from EDX, M1) at the $6c$ site and Ge+1.0Ga (M2 in Table\,\ref{StruDat})  at the $24k$ site, are shown in Table\,\ref{StruDat}. The atomic parameters are comparable with the available values of similar compositions in the literature.\cite{09Mel.An,2015Xu,06Joh.Si}          

\begin{figure}[H]
 \includegraphics[width=0.5\textwidth]{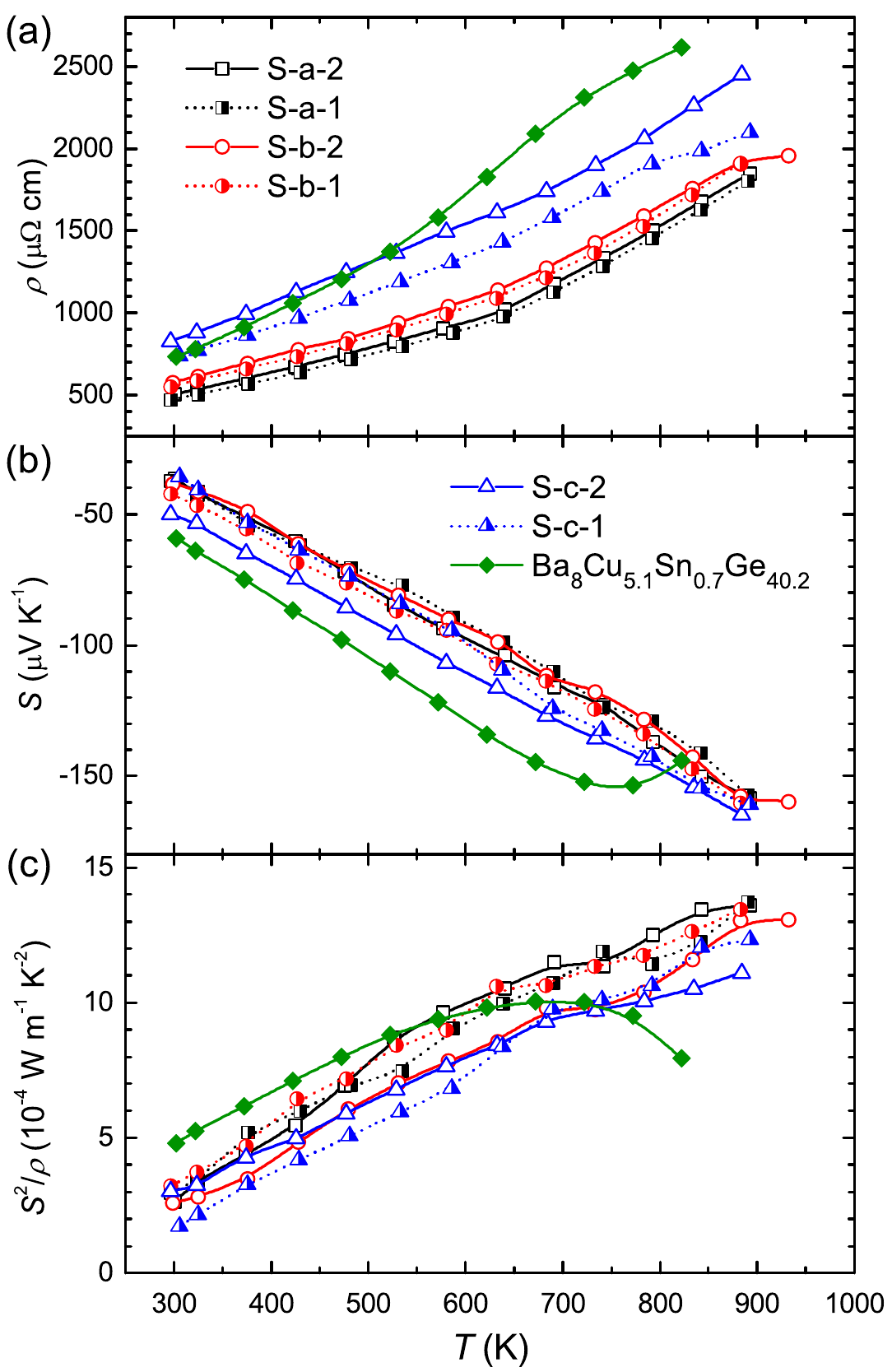}
 \caption{\label{Transports} 
(Color online) Temperature dependence of electrical resistivity $\rho$ (a), Seebeck coefficient $S$ (b), and power factor $S^2/rho$ (c) for S-p-i (p = a--c and i = 1, 2).  Data of Ba$_{8.0}$Cu$_{5.1}$Sn$_{0.7}$Ge$_{40.2}$ are plotted for comparison.\cite{2015Xu}  
}
 \end{figure}

\begin{figure*}[t]
\centering
 \includegraphics[width=1.0\textwidth]{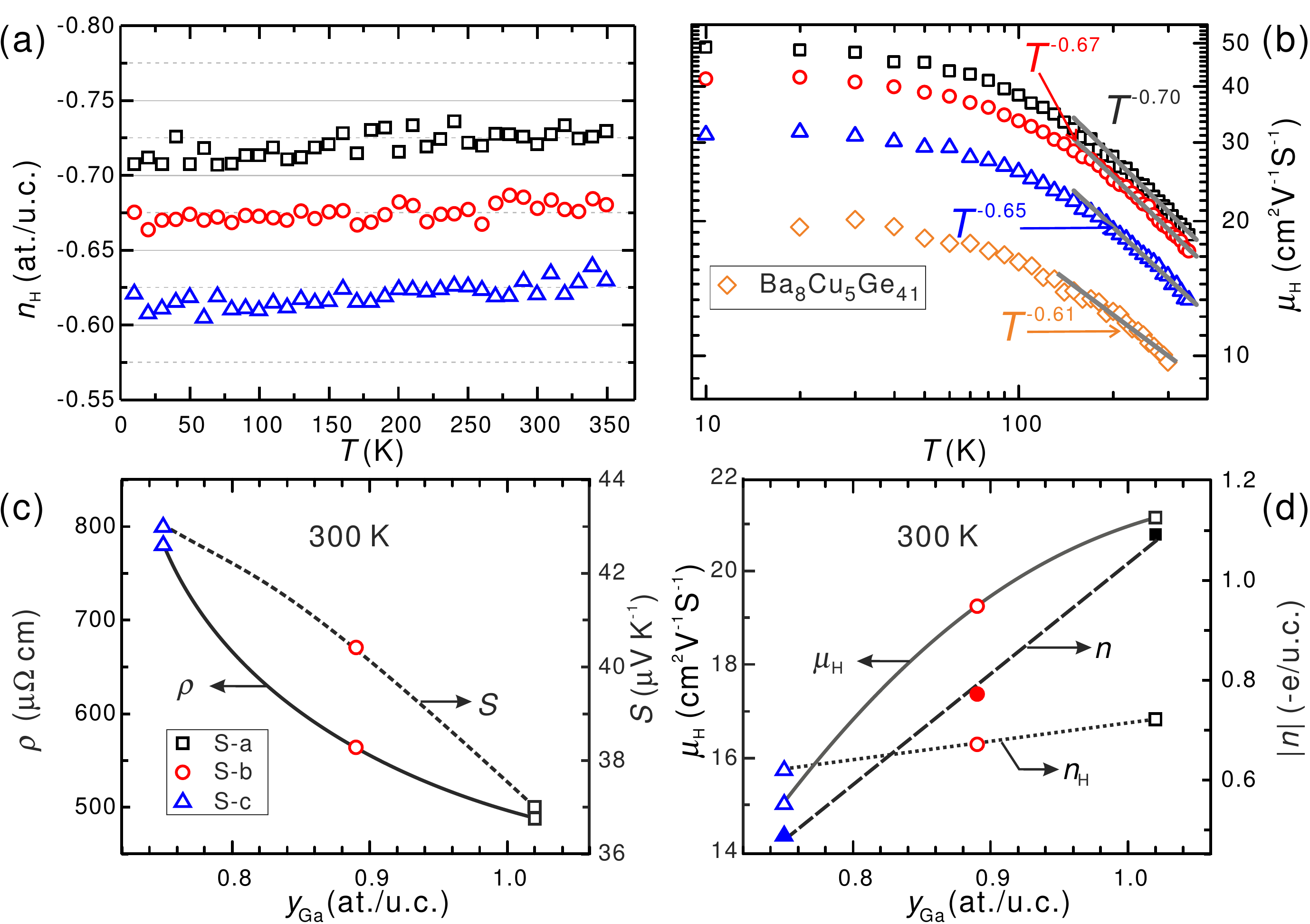}%
 \caption{\label{Charge-Mobility} (Color online) Absolute value of the charge carrier concentration $|n_{\rm H}|$ extracted from Hall effect measurements (a) and Hall mobility (b) versus temperature. The straight lines in (b) are fits with $\mu_H \propto T^\lambda$ to the data between 200 and 350 K. (c) Electrical resistivity and Seebeck coefficient at 300 K as function of the Ga content $y_{\rm Ga}$. (d) Theoretical carrier concentration $n$ derived from Eq.\,\ref{Zintl}, $|n_{\rm H}|$, and $\mu_{\rm H}$ versus $y_{\rm Ga}$ at 300 K. The lines in (c) and (d) are guides to the eyes.
}
 \end{figure*}

\begin{table}[h]
\small
  \caption{\ Selected interatomic distances (\AA, error bar = $\sim 0.0006$\AA) for the Ga-free single crystal of approximate composition Ba$_{8.0}$Cu$_{4.8}$Ge$_{41.2}$ [Ga0] and the Ga-containing single crystal of approximate composition Ba$_8$Cu$_{4.8}$Ga$_1$Ge$_{40.2}$ [Ga1], and their differences $\Delta d$ = $d_i$(Ga1)-$d_i$(Ga0), where $i$ denotes an interatomic distance such as Ba($2a$)--8Ge($16i$)}
  \label{StruDat-d}
  \begin{tabular*}{0.48\textwidth}{@{\extracolsep{\fill}}r@{--}l l l r}
    \hline
    $A$&n$B$ & Ga0     & Ga1 & $\Delta d$ \\
    \hline
 Ba($2a$) & 8Ge($16i$)  & 3.3932 & 3.3930 & -0.0002\\
         & 12Ge($24k$) & 3.5974 & 3.5935 & -0.0039\\
Ba(6$d$) &  8Ge($24k$) & 3.5668 & 3.5673 & 0.0005\\
         & 4Cu($4c$)   & 3.7802 & 3.7797 & -0.0005\\
         & 8Ge($16i$)  & 3.9773 & 3.9767 & -0.0006\\
         & 4Ge($24k$)  & 4.1232 & 4.1242 & 0.0010\\
Cu($6c$) & 4Ge($24k$)  & 2.4240 & 2.4272 & 0.0032\\
Ge($16i$) & Ge($16i$)  & 2.4731 & 2.4722 & -0.0009\\
         & 3Ge($24k$)  & 2.5031 & 2.5018 & -0.0013\\
Ge($24k$) & 1Cu($6c$)   & 2.424  & 2.4272 & 0.0032\\
         & 2Ge($16i$)  & 2.5031 & 2.5018 & -0.0013\\
         & 1Ge($24k$)   & 2.5613 & 2.5571 & -0.0042\\
    \hline
  \end{tabular*}
\end{table}

\begin{table*}
\small
  \caption{\ Structure data for the two single crystals with approximate compositions Ba$_{8.0}$Cu$_{4.8}$Ge$_{41.2}$ and Ba$_8$Cu$_{4.8}$Ga$_1$Ge$_{40.2}$ derived from X-ray single crystal refinements. Data collection: $2 \leq 2\Theta \leq 72.5$; 75 s/frame; Total number of frames: 210 frames in 5 sets; Mosaicity: $<0.43$; Derived space group (SG): $Pm\bar 3n$}
  \label{StruDat}
  \begin{tabular*}{\textwidth}{@{\extracolsep{\fill}}p{5.5cm} c c}
    \hline
    \bf Parameters/Compounds & \bf Ba$_{8.0}$Cu$_{4.8}$Ge$_{41.2}$ & \bf Ba$_8$Cu$_{4.8}$Ga$_1$Ge$_{40.2}$  \\
    \hline
    Composition (EDX, at/u.c.) & Ba$_{8.0}$Cu$_{5.0}$Ge$_{41.0}$ & \bf Ba$_8$Cu$_{\sim 4.6}$Ga$_{\sim 1}$Ge$_{\sim 40.4}$ \\
$a$ (nm)          & 1.06919(2) & 10.6906(2)\\
$a$ (nm), Ge standard & 1.06928(2) & 1.06975(1)\\
$\mu_{abs}$ (mm$^{-1}$) & 33.19 & 33.14\\
Reflections in refine & 515 (Fo) $\geq$ 4$\sigma$ (Fo) & 591 (Fo) $\geq$ 4$\sigma$ (Fo)\\  
Number of variables  & 18 & 20\\
$R_F^2=\sum|F_o^2-F_c^2|/\sum F_o^2$ & 0.019 & 0.021\\
$R_{in}$ &0.028 & 0.023\\
wR2 & 0.0489 & 0.0471\\
GOF & 1.407 & 1.671\\
Extinction coefficient & 0.00088(3) & 0.00095(4)\\
$R_e$ (highest peak; deepest hole) & 1.72;-1.20 & 1.58;-1.64 \\
Ba at $2a$ (0,0,0) U$_{eq}$ $10^2$ (nm$^2$) & 0.0095(1) & 0.0077(1) \\
U$_{ii}$,$i=1,2,3$ & 0.0095(1) & 0.0077(1)\\
Ba at $6d$ (1/4, 1/2, 0) U$_{eq}$ & 0.0340(1) & 0.0319(1) \\
U$_{11}$; U$_{jj}$,$j=2,3$ & 0.0225(2);0.0398(2) & 0.0195(2);0.0381(2)\\
M1 at $6c$ (1/4, 0, 1/2)  U$_{eq}$ & 0.0107(2) & 0.0088(2)\\
U$_{11}$; U$_{jj}$,$j=2,3$ & 0.0136(3);0.0092(2) & 0.0114(3);0.0075(2) \\
Ge at $16i$ ($x$,$x$,$x$)  U$_{eq}$  & 0.0083(1) & 0.0066(1)\\
$x$ & 0.18323(2) & 0.18324(2)\\
U$_{ii}$,$i=1,2,3$; U$_{23}$ = U$_{13}$ = U$_{12}$   & 0.0083(1);-0.00081(6)  & 0.0066(1);-0.00081(6)\\
M2 at $24k$ (0,$y$,$z$) U$_{eq}$  & 0.0097(1) & 0.00787(9)\\
$y$; $z$ & 0.31442(2);0.11978(2) &  0.31414(2);0.11960(2)\\
U$_{11}$; U$_{22}$ & 0.0095(1);0.0093(1) & 0.0076(1);0.0076(1)\\
U$_{33}$; U$_{23}$ & 0.0102(1);0.00101(8) & 0.0084(1);0.00093(8)\\
    \hline
  \end{tabular*}
\end{table*}

\subsection {Physical properties}
\subsubsection {Thermoelectric properties}

The temperature dependent electrical resistivity $\rho(T)$, Seebeck coefficient $S(T)$, and power factor $PF(T) = S^2/\rho$ for three pieces of the Ga-containing single crystal are shown in Fig.\,\ref{Transports}. $\rho(T)$ exhibits metal-like behavior for all samples and changes systematically from S-a to S-c (see also Fig.\,\ref{Charge-Mobility} (c)). $S(T)$ is negative and linearly dependent on temperature. The highest $PF$ value of 1.4 mW/mK$^2$ is reached in S-a at 900 K. This is about 30\% larger than the $PF$ of sample S-c-2 at the same temperature. The variations of these properties on temperature and composition can be qualitatively understood by the Zintl rule. As the framework atoms Cu and Ga consume 3 and 1 electron, respectively, of the electrons provided by Ba ($2\times 8$) for the bonding, the remaining number of electrons (in -e/u.c.) can be calculated by 
\begin{equation}
n=16-3\cdot x_{\rm Cu}-1\cdot y_{\rm Ga} \quad .
\label{Zintl}
\end{equation}
The negative sign of $S$ is thus related to the remaining nonbonded electrons (Table\,\ref{Tab:tab1}), and the increase of $\rho$ from S-a to S-c may,  at least in part, be due to an accompanying decrease of the charge carrier concentration $n$ (Fig.\,\ref{Charge-Mobility} (d)). Note that the decrease of the Cu content $x_{\rm Cu}$ associated with the increase of the Ga content $y_{\rm Ga}$ slows down the change of $n$ with composition, providing flexibility to finely tune the charge carrier concentration by composition.

The Hall effect analysis, however, shows that changes in the Hall mobility $\mu_{\rm H}$ dominate the change in electrical resistivity $\rho$. As expected from the Zintl rule, the experimentally determined $n_{\rm H}$ does indeed change with composition, but not as strongly as predicted by Eq.\,\ref{Zintl} (see Fig.\,\ref{Charge-Mobility} (d)). The Hall mobility, however, is strongly enhanced with increasing Ga content (Fig.\,\ref{Charge-Mobility} (d) and Table\,\ref{Tab:tab1}). At 300 K, for instance, $\mu_{\rm H}$ of S-a is 40\% larger than $\mu_{\rm H}$ of S-c, corresponding to almost the same relative reduction as that in $\rho$. In comparison, $n_{\rm H}$ increases by only 15\% (see Fig.\,\ref{Charge-Mobility} (d)).

\begin{figure*}
\centering
 \includegraphics[width=1.0\textwidth]{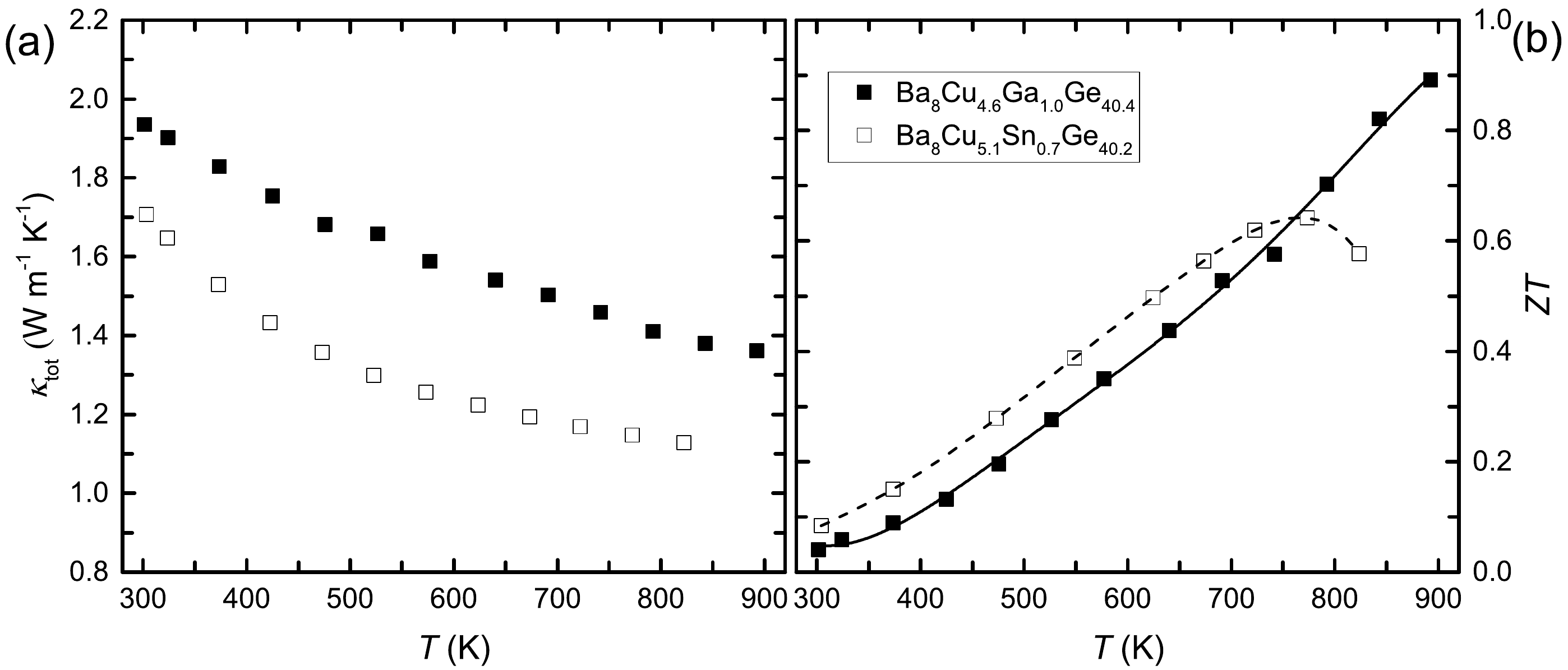}%
 \caption{\label{figureZT} Total thermal conductivity $\kappa_{\rm tot}$ (a) and figure of merit $ZT$ (b) as a function of temperature for the single crystal S-a. Data of Ba$_{8.0}$Cu$_{5.1}$Sn$_{0.7}$Ge$_{40.2}$ are plotted for comparison.\cite{2015Xu} Lines in (b) are guides to the eyes.   
}
 \end{figure*}

To understand the origin of the enhanced mobility, we analyzed the effective mass $m^*$ and the scattering parameter $\lambda$, which are related to the mobility by $\mu = e\cdot \tau /m^*$. Here, $\tau$ is the average relaxation time for all scattering processes, and $e$ is the electron charge. $m^*$ is estimated from $S(T)$ (Fig.\,\ref{Transports} (b))  by \cite{64Cu.Le}
\begin{equation}
S = \frac {2\pi ^2 k_B^2 m^*}{e \hbar^2(3n\pi ^2)^{2/3}}
\label{eqS1}
\end{equation}
and $\lambda$ is derived by fitting $\mu_{\rm H}(T)$ (Fig.\,\ref{Charge-Mobility} (b)) between 200 and 350 K with $\mu_H \propto T^\lambda$.  With increasing Ga content, both $m^*$ and $\lambda$ decrease slightly (Table \ref{Tab:tab1} and Fig.\,\ref{Charge-Mobility} (b)). The small decrease of $m^*$ (by only 5\% between S-a and S-c) can only partially account for the increase of $\mu_{\rm H}$. Therefore, we identify an increased relaxation time  $\tau$ as the main origin of the observed mobility enhancement. The $\lambda$ values between -0.65 and -0.70 are relatively close to -0.5, the value for alloy disorder scattering.\cite{05Ho.An} The increase of |$\lambda$| may be related to the vacancy filling, which leads to the decrease of alloy disorder scattering and thus the high mobility. To test this conjecture, we performed Hall effect  
measurement also on our Ga-free single crystal Ba$_8$Cu$_{4.8}$Ge$_{41.2}$, which has distinctly more vacancies than all Ga-containing crystals. Indeed, the Ga-free sample has the lowest |$\lambda$| ($\lambda$ is close to -0.5), and  the lowest mobility of all our crystals (Fig.\,\ref{Charge-Mobility} (b)). The mobility of 9.5 cm$^2$/Vs at 300 K for the Ga-free sample is even lower than the 11.9 cm$^2$/Vs for the  Sn-substituted single crystal Ba$_{8.0}$Cu$_{5.1}$Sn$_{0.7}$Ge$_{40.2}$ that has strongly distorted cages due to the large size of Sn.\cite{2015Xu}

Figures \,\ref{figureZT} (a) and (b) show the temperature dependent thermal conductivity $\kappa_{\rm tot}$  and the figure of merit $ZT$, respectively, for the sample S-a and for Ba$_{8.0}$Cu$_{5.1}$Sn$_{0.7}$Ge$_{40.2}$ for comparison.\cite{2015Xu} S-a has a higher thermal conductivity than the Sn substituted single crystal, which, however, is in part due to the higher electronic contribution $\kappa_e$. Below about 780 K, $ZT$ of our sample S-a is somewhat lower than that of the Sn-substituted crystal, which is mostly due to the lower $\kappa_{\rm tot}$  of the latter. At high temperatures, however, $ZT$ of our S-a crystal is much higher due to its high power factor (see Fig.\,\ref{Transports} (c)). The highest $ZT$ of 0.9 at  is achieved 900 K. 
As $ZT(T)$ is still not saturated at the highest temperature of our experiments, we anticipate even larger values at higher temperatures.

\subsubsection {Specific heat}

 \begin{figure}
 \includegraphics[width=0.5\textwidth]{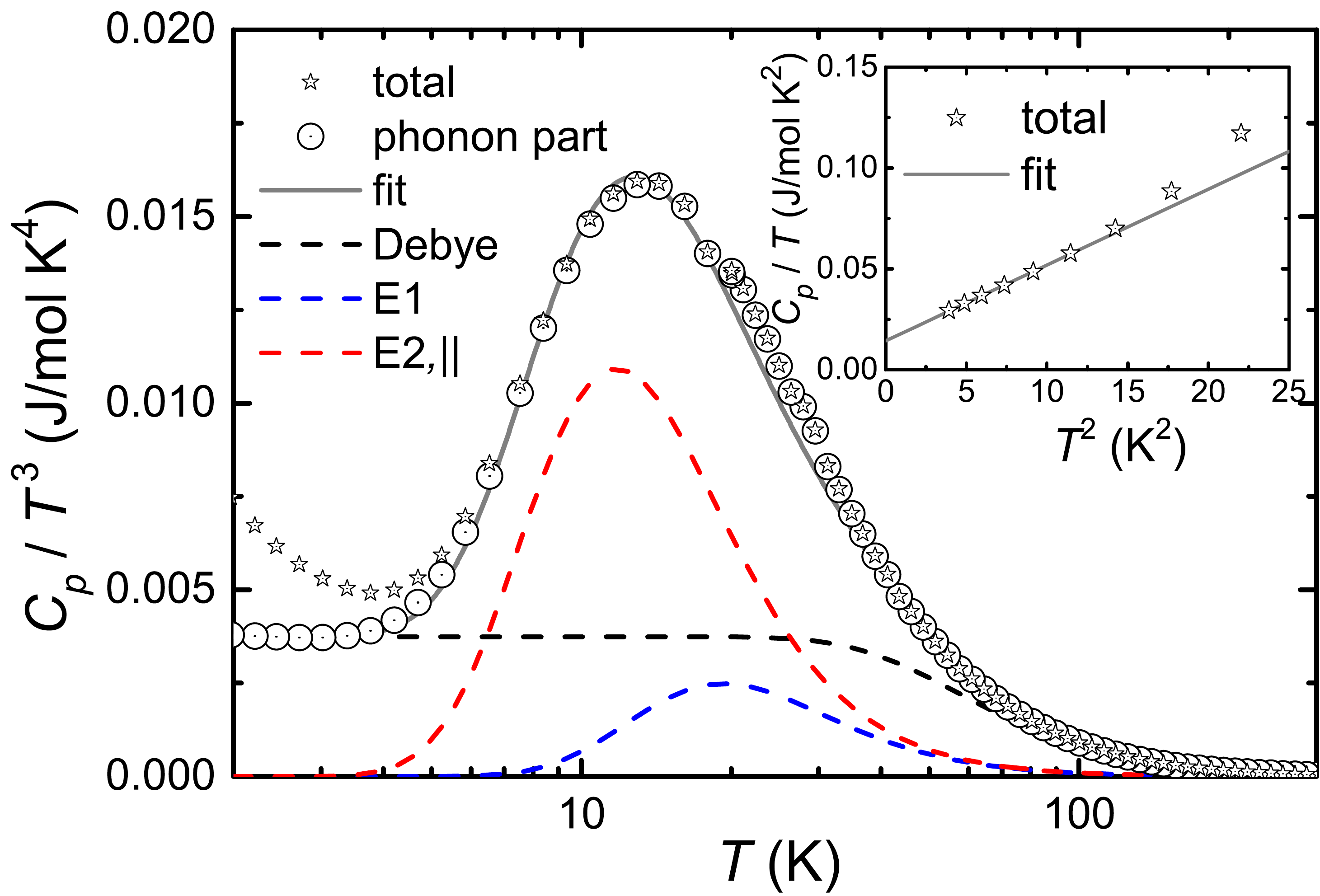}%
 \caption{\label{cp-low} (Color online) Temperature dependent specific heat $C_p$ of S-a, plotted as $C_p/T^3$ vs $T$. The full line is a fit of $C_p-\gamma T$, different contributions of $C_D$ (black) and $C_E$ (red and blue) are plotted as dashed lines. The inset shows data below 5 K plotted as $C_p/T$ vs $T^2$ and the fit according to Eq.\,\ref{Lowcp}. 
}
 \end{figure}

Specific heat $C_p$ data of S-a are shown in Fig.\ref{cp-low}. Below 4 K the standard description is assumed, i.e.,
\begin{equation}
C_p/T=\gamma+\beta T^2\quad,
\label{Lowcp}
\end{equation}
where $\gamma$ is the Sommerfeld coefficient of the electronic contribution and $\beta$ the low-temperature coefficient of the lattice contribution. The fit (Fig.\ref{cp-low}, insert) yields $\gamma = 14.4$ mJ/(molK$^2$) and $\beta = 3.75$ mJ/(molK$^4$). Using  $\theta_D=(12\pi^4RN/5\beta)^{1/3}$, where $R$ is the gas constant and $N$ the number of atoms per u.c., and treating the guest atoms as independent Einstein oscillators and the framework atoms as a Debye solid, i.e., $N= N_D=46$, we obtained $\theta_D=287$ K. Both $\gamma$ and $\theta_D$ are in good agreement with values derived for Ba$_8$Cu$_{5.3}$Ge$_{40.7}$ ($\gamma = 11.8$ mJ/molK$^2$ and $\theta_D=289$ K).\cite{12Xu.He}

To model the data in the entire temperature range (Fig.\,\ref{cp-low} main panel) we used
\begin{equation}
C_p=\gamma T+C_D+C_E \quad ,
\label{wholeTcp}
\end{equation}
where $C_D$ and $C_E$ are the Debye and the Einstein contribution, respectively, namely
\begin{equation}
C_D=9N_DR\left(\frac{T}{\theta_D}\right)^3\int_0^{\theta/T}\frac{x^4\ue^x}{(\ue^x-1)^2}\,\ud x\quad,
\end{equation} 
\noindent with $x=\hslash\omega/(k_BT)$ and the phonon-angular frequency $\omega$, and 
\begin{equation}
C_E=\sum_{i=1}^Np_iN_{Ei}R\left(\frac{\theta_{Ei}}{T}\right)^2\frac{\ue^{\theta_{Ei}/T}}{(\ue^{\theta_{Ei}/T}-1)^2}\quad,
\end{equation} 
\noindent where $p_i$ is the number of degrees of freedom, $N_{Ei}$ the number of Einstein oscillators, and $\theta_{Ei}$ the Einstein temperature of the $i$th vibrational mode. 

With the constraints for type-I clathrates given in Table\,\ref{Tab:tab2},\cite{10Chr.Jo,11Yan.Gr} the data are well described with two Einstein temperatures, representing vibrations in two perpendicular directions for Ba at the 6$d$ site, and one Einstein temperature for Ba at the 2$a$ site (see Table\,\ref{Tab:tab2}).

\begin{table}[h]
\small
  \caption{\ Constraints evaluated from the structure characteristic of type-I clathrates,\cite{10Chr.Jo,11Yan.Gr} used in the fit of $C_p(T)$ with Eq.\,\ref{wholeTcp}, and corresponding results. The subscripts number 1 and 2 denote atoms at the $2a$ site (dodecahedral cages) and the $6d$ site (tetrakaidecahedral cages), respectively. $\parallel$ and $\perp$ represent vibration directions of the atoms at the $6d$ site parallel and perpendicular to the 6-atom-ring planes of the tetrakaidecahedra.}
  \label{Tab:tab2}
  \begin{tabular*}{0.48\textwidth}{@{\extracolsep{\fill}} l l}
    \hline
Constraints & $\theta_{E1}>\theta_{E2}$, $\theta_{E2}^\perp>\theta_{E2}^\parallel$\\
            & $N_{E1}=2$, $N_{E2}=6$, $p_1=3$, $p_2^\parallel=2$, $p_2^\perp=1$\\
Results & $\theta_{E1}=\theta_{E2}^\perp =95$ K, $\theta_{E2}^\parallel = 58$ K\\
    \hline
  \end{tabular*}
\end{table}

\section{Conclusions}

Our detailed crystal structure and thermoelectric property investigation of type-I Ba$_8$(Cu,Ga,Ge,$\Box$)$_{46}$ clathrate single crystals unambiguously revealed that vacancies, present at the $6c$ site in Ba$_8$Cu$_{4.8}$Ge$_{41.2}$, are successively filled upon Ga substitution. This was revealed by an X-ray single crystal diffraction study, demonstrating an interatomic distance elongation for the Cu(6$c$)-Ge(24$k$) distance and a shrinking for both the Ge(24$k$)-Ge(24$k$) distance and the diameter of the small cages in Ba$_8$Cu$_{4.8}$Ga$_1$Ge$_{40.2}$. The vacancy filling removes local disorder and leads to an increased charge carrier mobility and thus to enhanced thermoelectric performance. In view of the narrow composition range (Cu: $\sim$4.6 to $\sim$4.9 at./u.c., Ga: $\sim$1.0 to $\sim$0.7 at./u.c.) in our single crystal, the size of the enhancement is surprisingly large. The highest figure of merit $ZT=0.9$ at 900 K was achieved for a single crystal with an approximate composition Ba$_8$Cu$_{4.6}$Ga$_{1.0}$Ge$_{40.4}$. This value, that has still not reached saturation at the highest temperature of our measurements, is to date one of the largest in transition metal element-containing clathrates. We conclude that reducing the vacancy content in type-I clathrates is an important design strategy to optimize their thermoelectric performance.   

\section*{Conflict of interest}
There are no conflicts to declare.

\section*{Acknowledgments}

We thank M. Waas for  SEM/EDX measurements. This work was supported by the DFG project SPP1386, the FWF projects TRP 176-N22 and I2535-N27, the Christian Doppler Laboratory for Thermoelectricity, and the European C-MAC.

\balance



\providecommand*{\mcitethebibliography}{\thebibliography}
\csname @ifundefined\endcsname{endmcitethebibliography}
{\let\endmcitethebibliography\endthebibliography}{}

\end{document}